\documentclass[a4paper,12pt]{article}
\usepackage{epsfig,array,amscd}
\usepackage{amsmath,amssymb}
\def\bf{\bfseries}

\title{\bf CAN SMALL FREE METHANE CLUSTERS EXHIBIT PHASE-LIKE TRANSITIONS?}
\author{E. Daykova and A. Proykova ${^\dag}$}
\date{Department of Atomic  Physics,\\
Faculty of Physics,Sofia University,\\
Sofia-1126, 5 J. Bourchier Blvd.\\
${^\dag}$anap@phys.uni-sofia.bg}

\begin{document}
\sloppy
\frenchspacing

\maketitle

\begin{abstract}
Low-temperature ($<$ 60K) phase changes of free
clusters containing 50, 137, and 229 $CH_4$ molecules
have been observed in 
isoenergetic Molecular Dynamics computations. Bulk solid methane
exhibits structural phase transformation at 20.4 K. However,
 clusters of 50 molecules already melt
at $\sim$ 25 K, which makes the observation of solid-solid transformations
rather challenging.  
\end{abstract}

\section{Introduction}

Small free clusters of molecules with very high symmetry (octahedral) have
been found to resemble bulk of the same substance even in the case
of small number of ingredients (30 to 80) \cite{A-B97}. The number
of particles (molecules or/and atoms) that mimics bulk properties like
phase transitions is strongly
dependent on the range of the potential. 
The effect of decreasing the range of the potential is
 to destabilize strained structures \cite{DWB}.
 In the case of long-range potentials 
(Coulomb \cite{A-B97}) one detects structures, determined by the
co-operative (collective) interaction - the system is expected to
 transform easier than in the case of short-range interactions. On
the other side, the surface of free clusters plays much more important role
 for long-range potentials. Various parameters of the potential, for instance
 electric charge value, govern the cluster response  
to external changes. An example is the melting temperature increase
 for higher charge values
  \cite{Pisov-CPC}. 

In the present work we study clusters consisting of less symmetrical molecules
(tetrahedral) in order to check how the lower symmetry of the ingredients
influences cluster's behavior at different temperatures. The interacting
 potential contains a long-range term (Coulomb), which makes it necessary to
study the size influence - clusters of 50, 137, 229 methane molecules
have been simulated with the help of the isoenergetic Molecular Dynamics
method described in the next section. Simulations of methane clusters have been
 performed more than two decades ago \cite{Ruth-B}. 
They showed that the lower limit
of the classic approach is $\sim$10K. 
In those days the computers were less fast and the researchers
 tended to go for cheap
rather than reliable potentials. Nowadays more sophisticated potentials
are available and better results could be obtained \cite{Allen}. This
justifies re-visiting the problem of how the cluster size influences
the structural phase changes. Another intriguing question is how a free
cluster, e.g. a cluster at zero external pressure, behaves when the
temperature is rather close to  the ultimate low
 limit of the classic mechanics calculations.\\

Solid (bulk)
 methane exhibits several crystalline phases based mainly on the difference
 in the orientation of molecules in the unit cell.
Infrared investigation of solid methane performed by Bini
and co-workers \cite{Bini} let him propose a phase diagram,
which has been additionally elucidate by Nakahata {\it et al.} with
a help of optical and X-ray 
diffraction studies \cite{Nakahata}.
The liquid methane freezes at 90.6 K and  $p = 0.1 $ MPa \cite{Amey} 
  into 
a {\it fcc} orientational disordered phase I \cite{Bell}, characterized with
rotational diffusion of the molecules.
A transition to the orientational ordered 
phase II occurs at 20.4 K \cite{Clusius29} at zero pressure.
 The phase II is the James-Keenan cubic 
structure, in which one out of four molecules ($O_h$) is undergoing 
a weakly hindered rotation, while the other three molecules ($D_{2d}$) have an 
anti-ferro order \cite{W.Press}.

The state of orientational disorder takes place for the systems of spherically-
symmetric molecules ($CH_4$) with a symmetry lower than the site in 
the crystal structure  \cite{Bell}. At cooling such systems could undergo phase
 transitions to a partially-ordered or fully ordered state. Hindered rotations 
occur because of weakly angular-dependent intermolecular forces and a large 
rotational energy as it is in methane \cite{W.Press}. 

Is such a behavior inherent for small systems as well? This is the question
we study in the current work.

\section{The model} 
Due to their spherical symmetry methane molecules resemble in many aspects the 
noble gases (Ar, Kr, Xe). The intra-molecular frequencies 
($\sim 4 \times 10^{13}, \sim 9 \times 10^{13}$ Hz) are 
one order of magnitude higher than
 the frequency of the intermolecular vibrations. That is why, we 
consider the molecules as rigid bodies (C-H distance b=1.094 {\it \AA}).
 Suitable intermolecular potentials for regarding 
molecular space orientations could be the Lenard-Jones potential 
\cite{Allen} or  a more sophisticated 
3-body RMK potential \cite{RMK}. In both approaches complicated  procedures 
had been followed to obtain parameters' values. Needless to say, these 
parameters are not entirely satisfactory.
In our study, the total potential is  a sum of  pair atom-atom potentials
$U_{pw}(i,j)$:

\begin{eqnarray}
\label{pot}
U_{pw}(i, j) &=&
 \sum\limits_{\alpha, \beta = 1}^5
\Biggl \lbrack 4  \, \epsilon_{\alpha \beta} \biggl  \lbrack \biggl(
\frac{\sigma_{\alpha \beta}}{r_{ij}^{\alpha \beta}}\biggr)^{12}
-
\biggl(
\frac{\sigma_{\alpha \beta}}{r_{ij}^{\alpha \beta}}
\biggr)^6 \biggr \rbrack \nonumber 
+\frac{q_{i \alpha} q_{j \beta}}{4 \pi \epsilon_0 r_{ij}^{\alpha \beta}}
\Biggr
\rbrack \\
U_{pot} &=& \sum\limits_{i,j = 1 \,(i < j)}^N {U_{pw}(i, j)} \nonumber
\end{eqnarray}
where $\it \alpha$ , $\it \beta$ denote either a carbon or a hydrogen atom;
$r_{i,j}$ is the distance between the $i$-th and the $j$-th atom;
$q_{i,\alpha}$ is either $q_C = -0.572$ or $q_H= 0.143$ in
elementary charge units.\\
The parameters have been obtained in a way to fit 
experimental data. For instance, the charge and the 
bond-length are determined from the octopole moment value. 
$\it \sigma_{\alpha\beta}$ in {\it \AA}, $\it \epsilon_{\alpha\beta}$ in 
{\it meV} are \cite{Allen}:\\
$\sigma_{CC}=3,35$ , $\sigma_{HH}=2,81$ , $\sigma_{CH}=3.08$ ;\\
$\epsilon_{CC}=4.257$ , $\epsilon_{HH}=0.715$ , $\epsilon_{CH}= 1,7447$. \\
Two starting configurations with randomly oriented 
molecules have been designed: 
1) a {\it bcc} lattice, which resembles the most 
spherically-symmetric shape of a cluster and 2) {\it fcc},
which is a possible lattice for a bulk methane.
We execute the velocity Verlet algorithm \cite{Allen,Verlet} for numerical integration of 
classic equations of motion in a special micro-canonical ensemble - N, E, p=0 
(free clusters) at temperatures above 10{\it K} to avoid quantum effects. The 
integration step of $dt=1fs$ guarantees  an accuracy
 greater than 0.003\% of the energy conservation 
for  runs up to $\sim $ 0.5{\it ns}; the records are taken
every  100 {\it fs}.
\section{Results}
Liquid- and solid-like phases can be distinguished with the help of a
specially constructed formula, which is an extension of the 
Lindemann index $\delta_{lin}$ \cite{Lind}:
\begin{equation}
\label{lin}
\delta_{lin} = \frac{2}{N(N-1)} \sum_{i,j(i<j)=1}^{N} \frac{\sqrt{\langle r_{ij}^2 \rangle - \langle r_{ij} \rangle ^2}}{\langle r_{ij} \rangle}
\end{equation}
\noindent where $r_{i,j}$ is the distance between molecular centers of mass.
 In bulk, the Lindemann index is computed on the basis of the mean  deviation
of the $i$-th atom from its ideal lattice position
and $\delta_{lin}<0.1$ indicates a solid state.
 In a free cluster the surface plays a destabilizing role
 and the cluster is solid if $\delta_{lin}<0.08$ \cite{Pisov-JCP}.

\begin{figure}
 \includegraphics[width=0.48\textwidth]{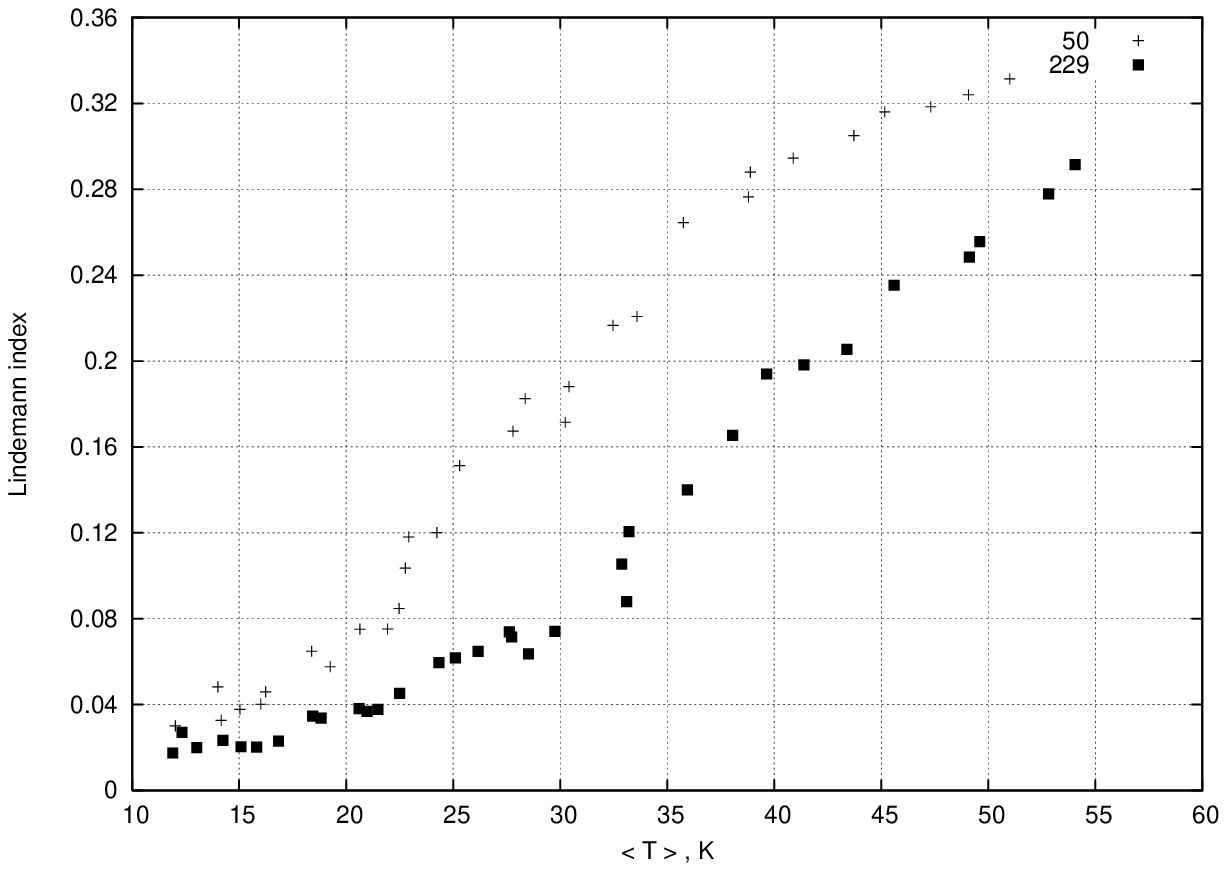}
 \hfill
 \includegraphics[width=0.48\textwidth]{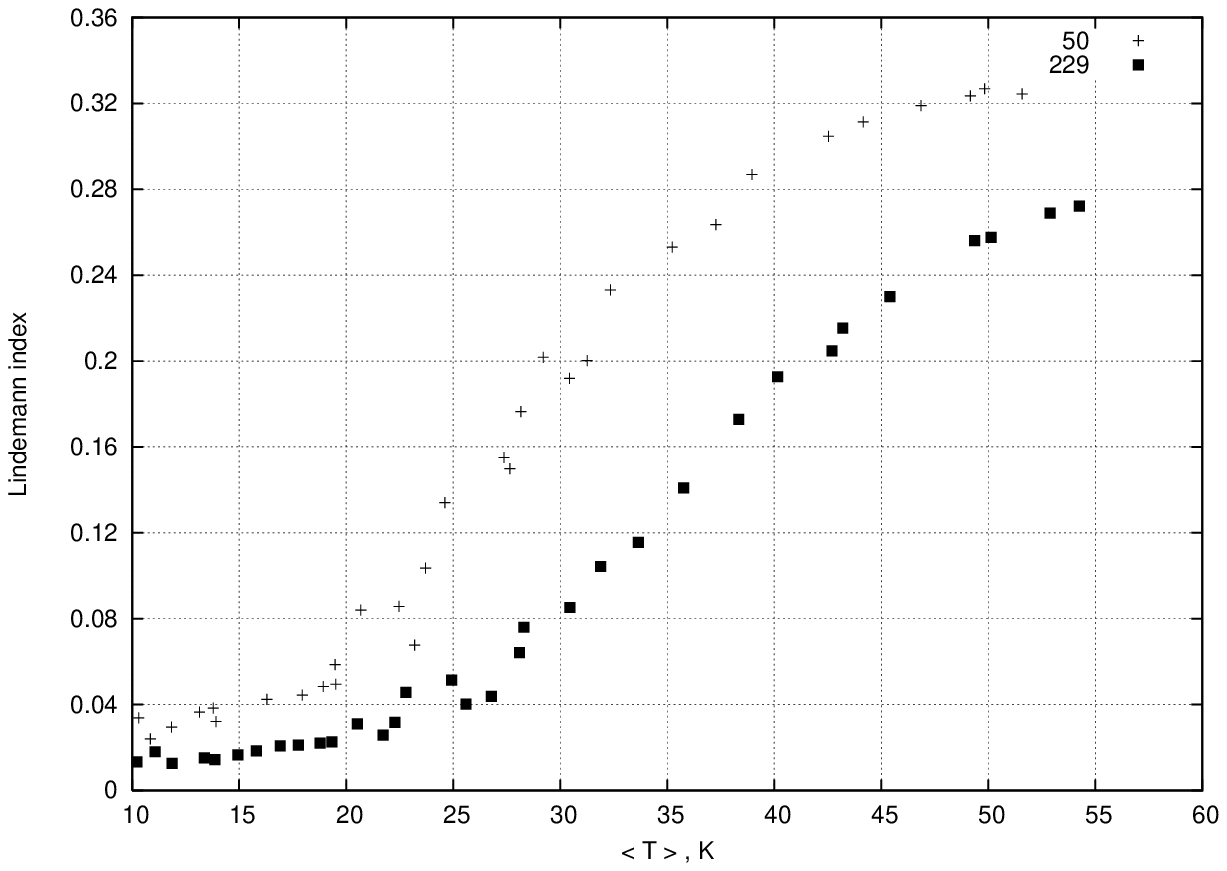}
 \\
 \parbox[t]{0.48\textwidth}{\caption{Lindemann index for clusters of 50 and 229 molecules; a {\it bcc}-starting configuration.}\label{lin-b}}
 \hfill
 \parbox[t]{0.48\textwidth}{\caption{Lindemann index for clusters of 50 and 229 molecules; a {\it fcc}-starting configuration.}\label{lin-f}}
\end{figure}

\begin{figure}
 \includegraphics[width=0.48\textwidth]{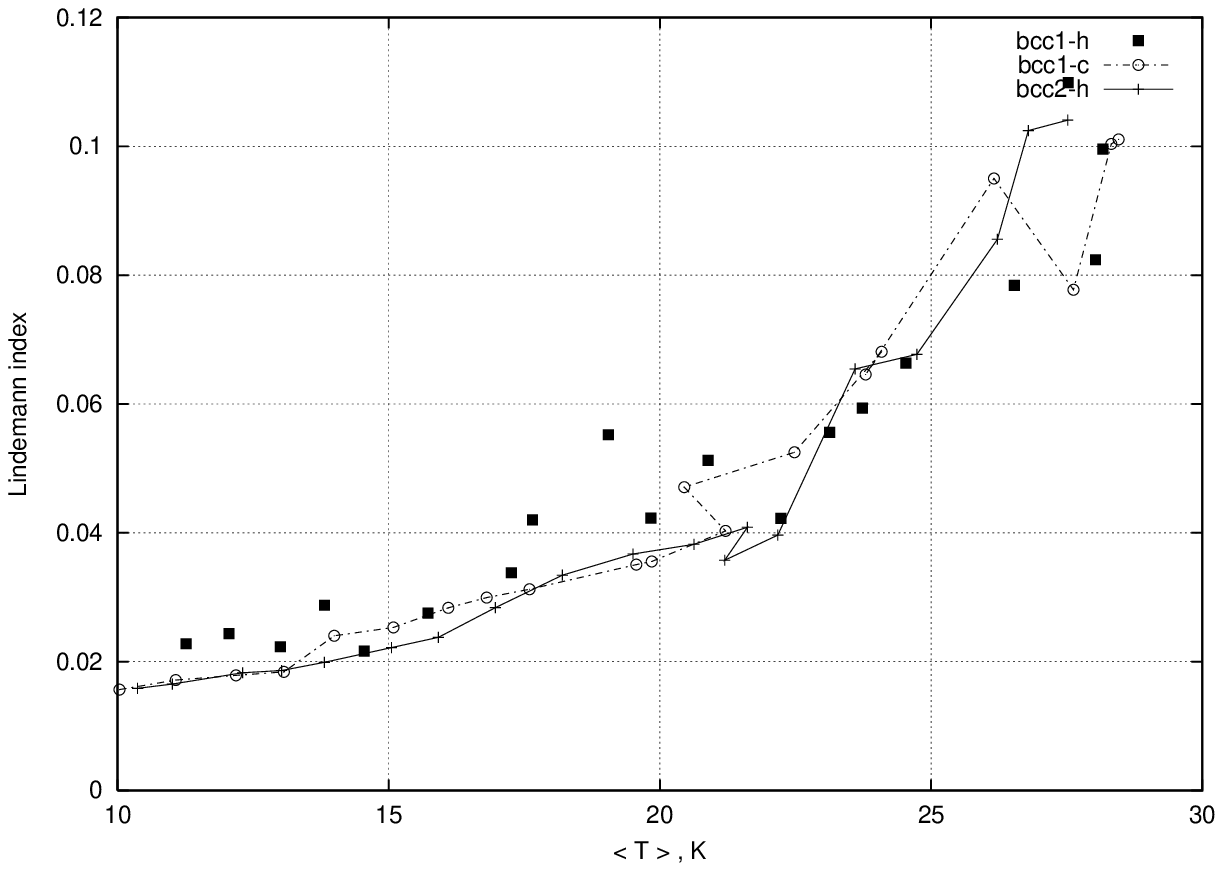}
 \hfill
 \includegraphics[width=0.48\textwidth]{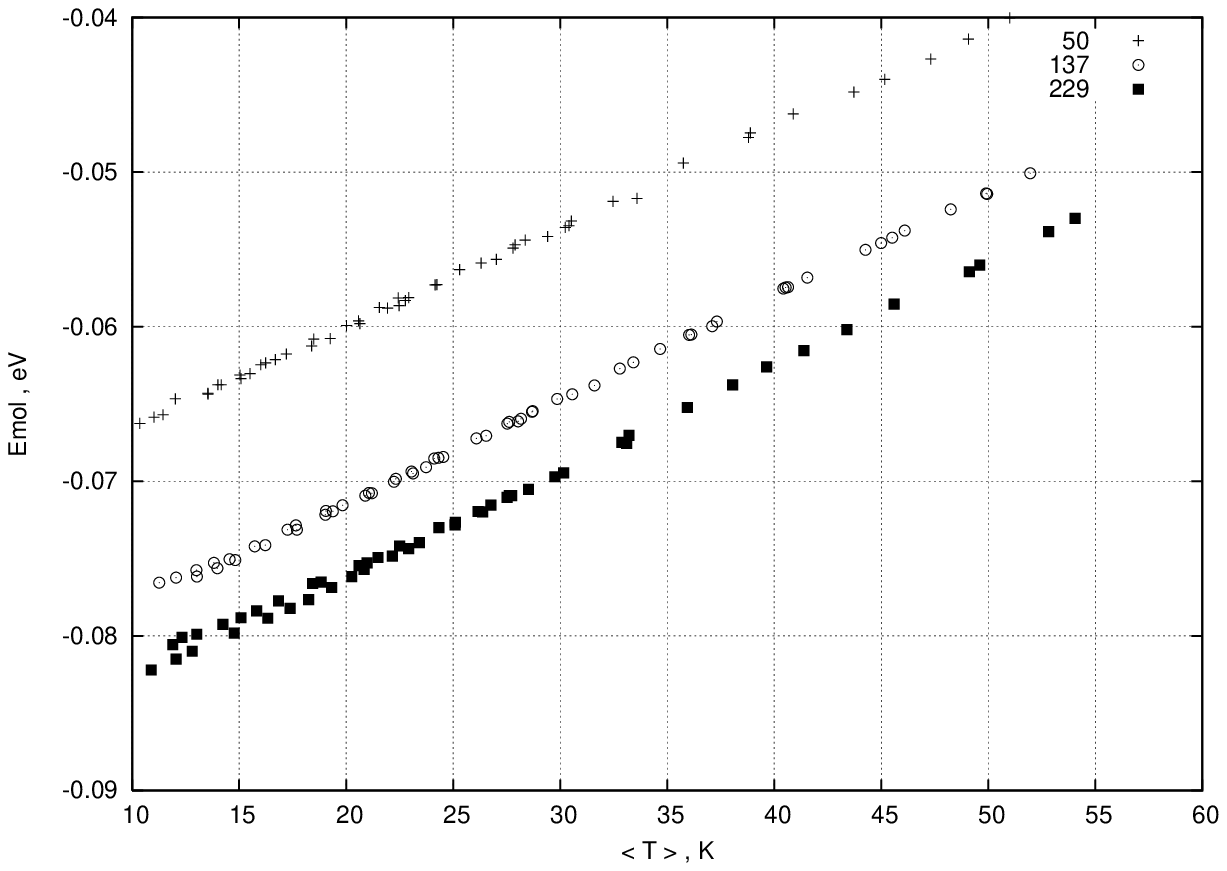}
 \\
 \parbox[t]{0.48\textwidth}{\caption{Lindemann index for heating-cooling-heating of 137 molecule cluster; a {\it bcc}-starting configuration.}\label{137lin}}
  \hfill
 \parbox[t]{0.48\textwidth}{\caption{No hysteresis is observed on  'cooling-heating' for the two starting configurations; only the case of a {\it bcc}-starting configuration is shown.}\label{nohyst}}
\end{figure}
\begin{figure}
 \includegraphics[width=0.48\textwidth]{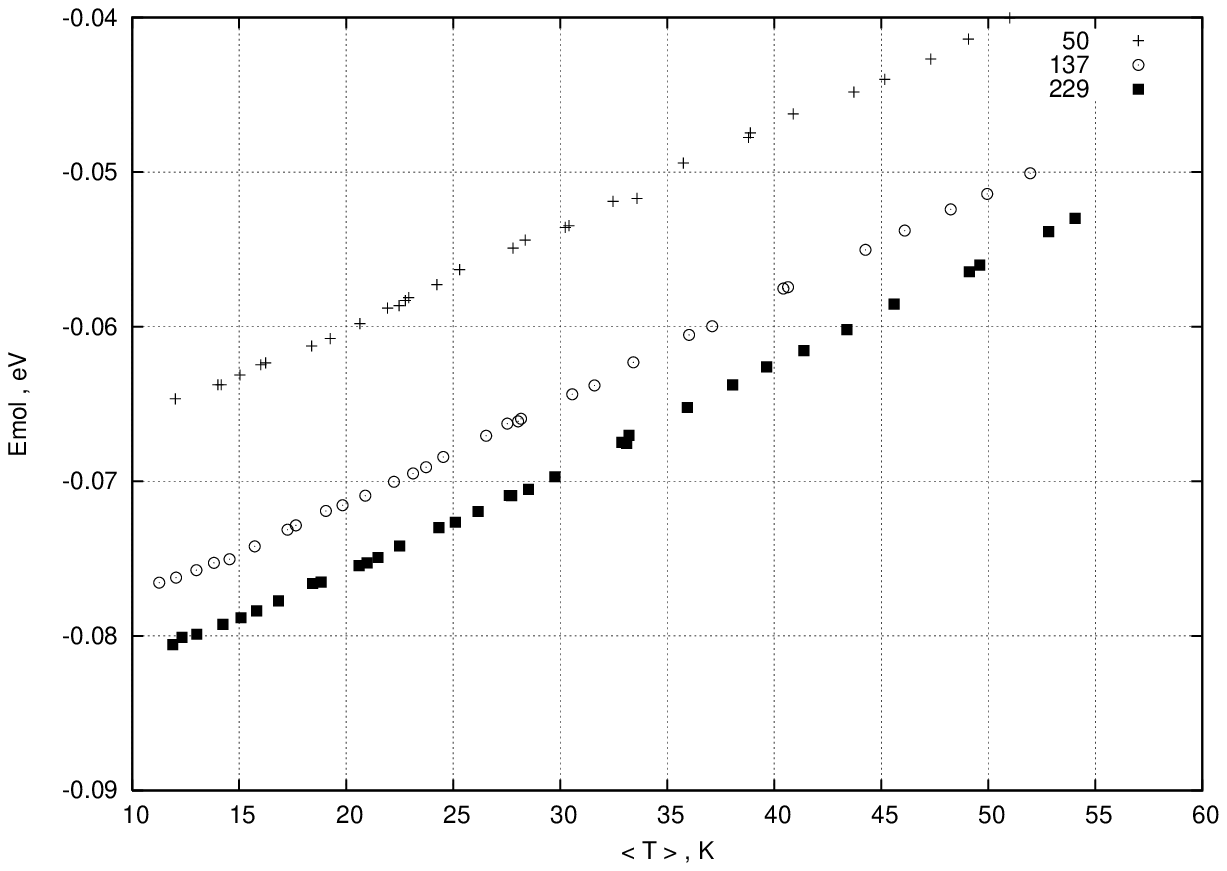}
 \hfill
 \includegraphics[width=0.48\textwidth]{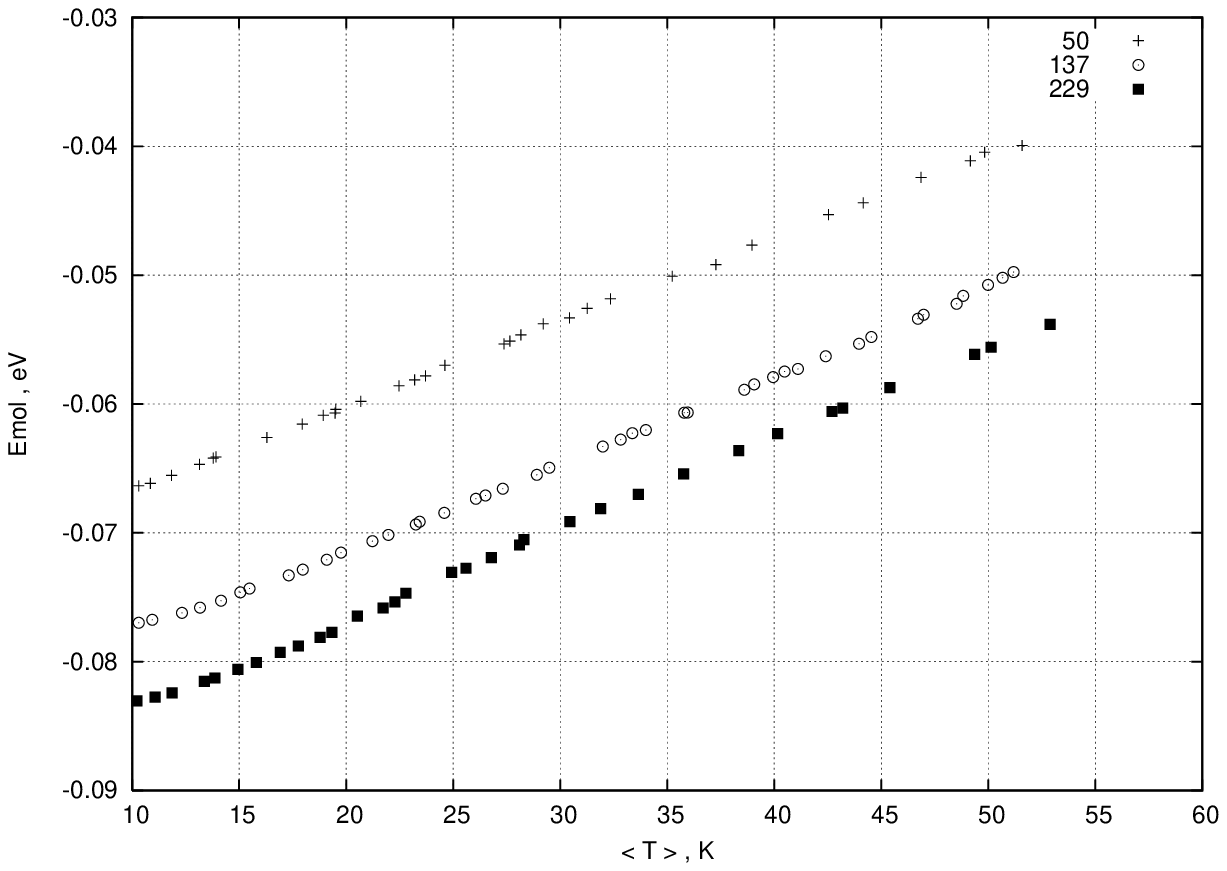}
 \\
 \parbox[t]{0.48\textwidth}{\caption{Caloric curves at heating of  clusters containing 50, 137, and 229 molecules, a {\it bcc}-starting configuration.}\label{cal-b}}
 \hfill
 \parbox[t]{0.48\textwidth}{\caption{Caloric curves for the same cluster sizes with a {\it fcc}-starting configuration.}\label{cal-f}}
\end{figure}

The Lindemann index, plotted in the Figs.(\ref{lin-b} - \ref{137lin}),
shows three distinguishable phases : two solid-like phases below 30{\it K} 
and a liquid-like phase above 30{\it K} for all three sizes and both starting 
configurations.\\
The caloric curves (total energy per particle as a function of temperature) are
given in the Fig.\ref{cal-b} and Fig.\ref{cal-f}.  Although
the caloric curves are too smooth (a change of the slope is hardly seen,
no more than 6\%),
 we observe a region of frustration in the temperature interval (18,25) {\it K}.
In our computations the average temperature of the system is computed from
its average kinetic energy: $<E_{kin}> = 3/2 k_B T$, with $k_B$ - the Boltzmann
constant. Analyzing the trajectories at different total energies, we
see that the system temperature jumps unevenly in the above interval. This
is an indication for changes much more clearly seen with the
help of the Lindemann index.
There is no hysteresis 
on 'heating-cooling' of the system - Fig.\ref{nohyst}.

The different phase structures are distinguished on the basis of their radial 
distribution functions - Eq. \ref{gr}:
 the radial distribution of molecular centers of mass shows the structure of the cluster at a  specific temperature, 
while the atom-atom radial distribution reveals orientational order (disorder) of the molecules. 

\begin{equation} \label{gr}
g(r) =
\frac{V}{N^2}\langle \sum\limits_{i=1}^{N} \sum\limits_{j\neq i}^{}
\delta(r-r_{ij}) \rangle
\end{equation}

\begin{figure}
 \includegraphics[width=0.48\textwidth]{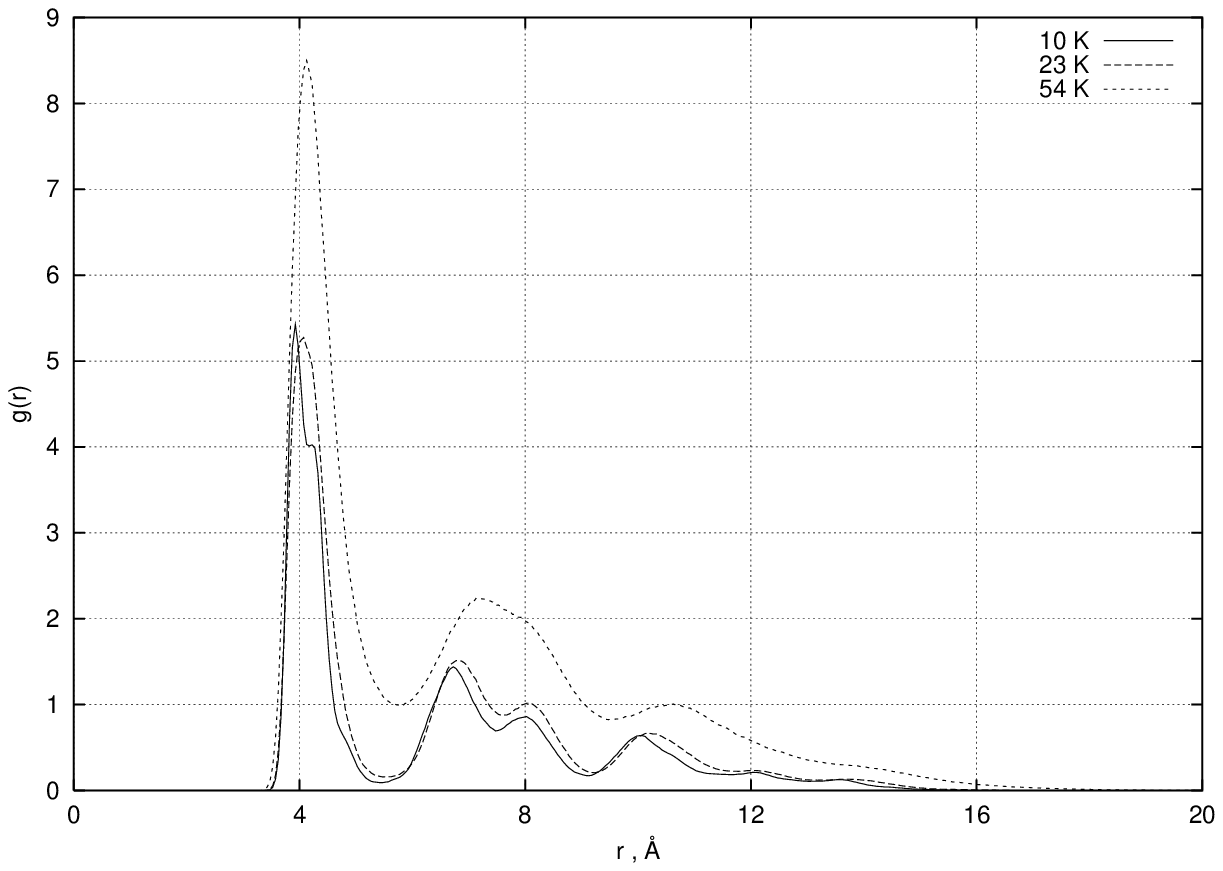}
 \hfill
 \includegraphics[width=0.48\textwidth]{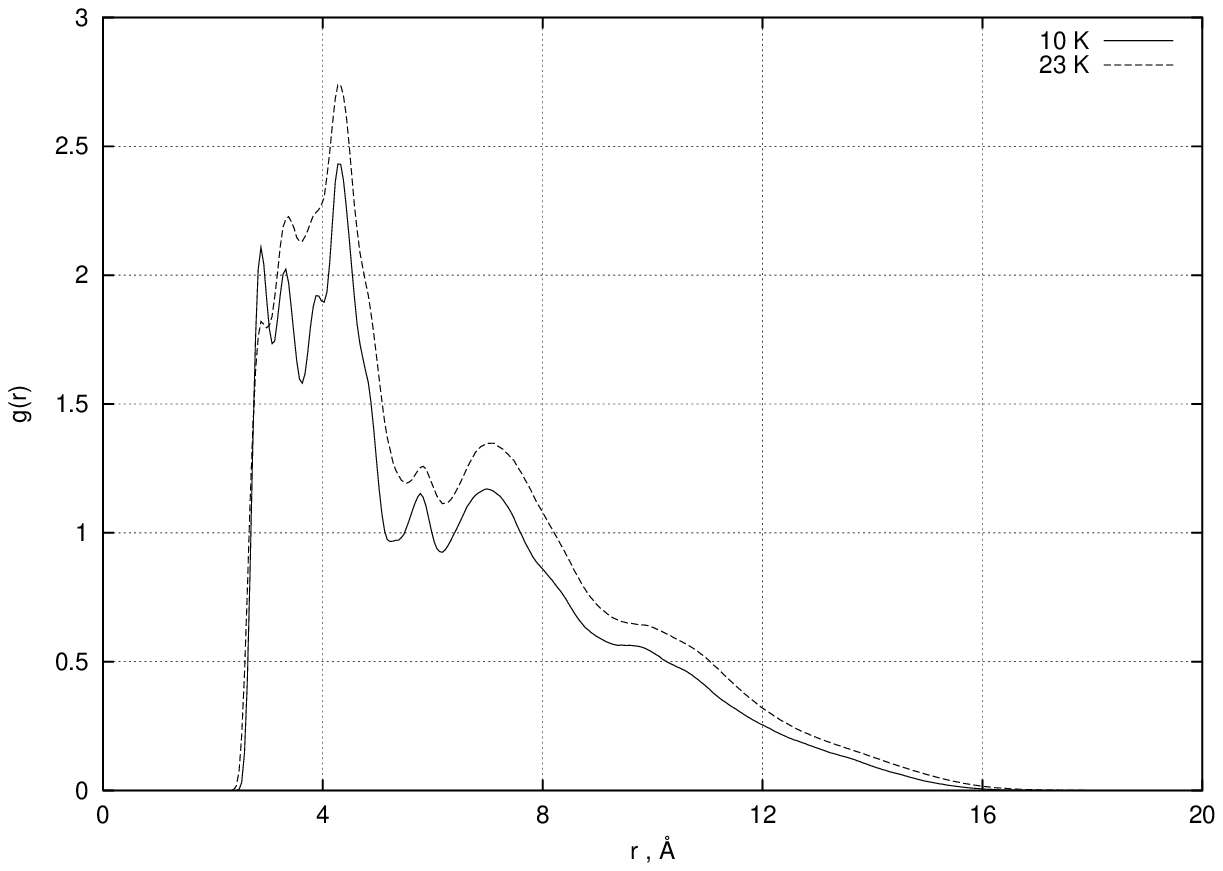}
 \\
 \parbox{0.48\textwidth}{\caption{Radial distribution of the molecular centers
 of mass for a 50 $CH_4$ cluster; a {\it fcc}-starting configuration.}\label{gr-mm}}
 \hfill
 \parbox{0.48\textwidth}{\caption{Atom-atom radial distribution for 50 $CH_4$ cluster, {\it fcc}-start.}\label{gr-at}}
\end{figure}

Fig.\ref{gr-mm} shows the existence of a solid-like structure below 30 {\it K} 
and the absence of that structure above 30 {\it K}. The volume of the cluster
 enlarges
at heating as well. Fig.\ref{gr-at} shows typical 
distribution of the atoms for the oriented (solid line in the figure)
and disoriented (dashed line in the figure) 
solid-like phases.
The first peak of the distribution correspond to the 
H-H neighbors, the second peak corresponds to the C-H neighbors, and the 
third one is for the C-C neighbors. The same maximum of the H-H and 
C-H peaks indicates a unique mutual orientation of the molecules, e.g
the cluster is orientationally ordered. The phase transition at $\sim$ 20K
causes disorientation of the molecules as it is seen in the 
Fig.\ref{gr-at} - the dotted curve.\\
\begin{figure}
 \includegraphics[width=0.48\textwidth]{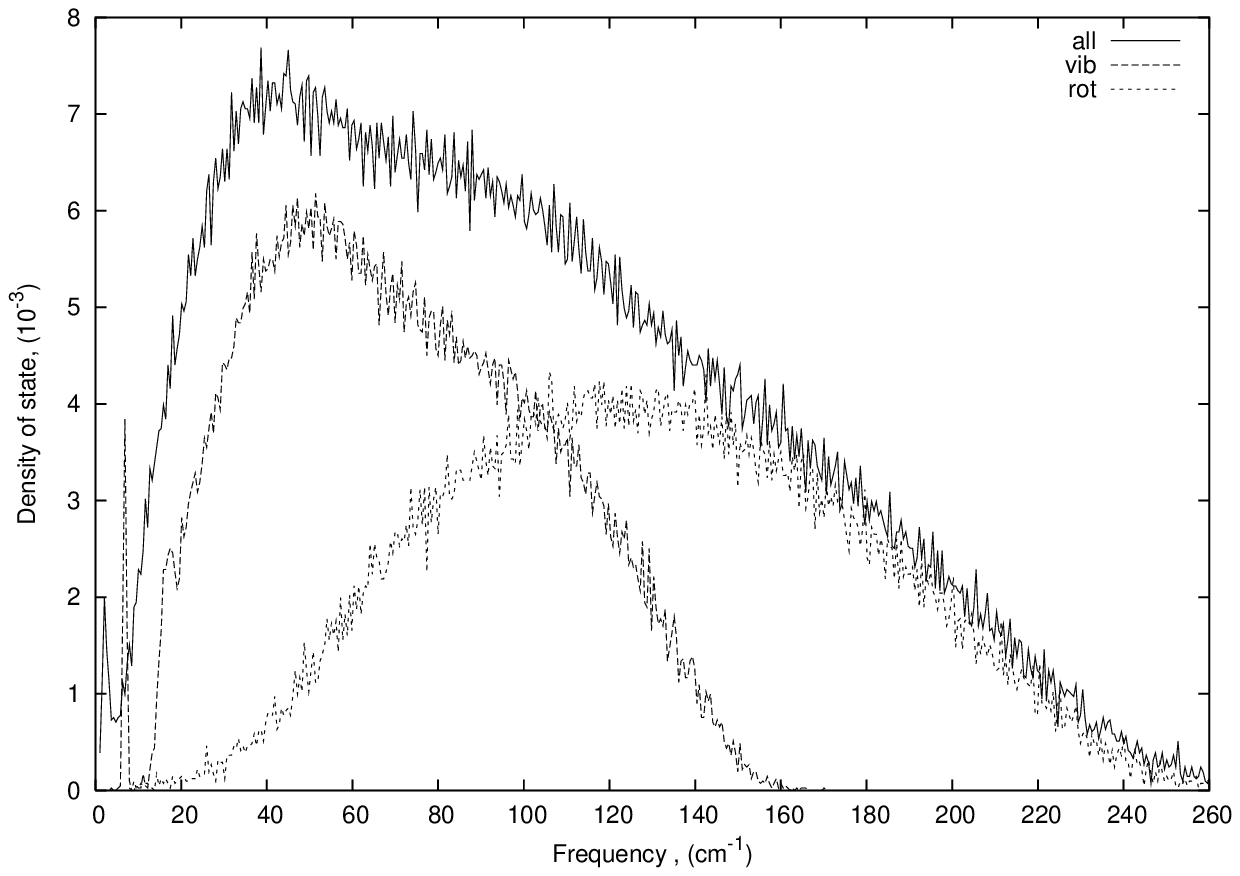}
 \hfill
 \includegraphics[width=0.48\textwidth]{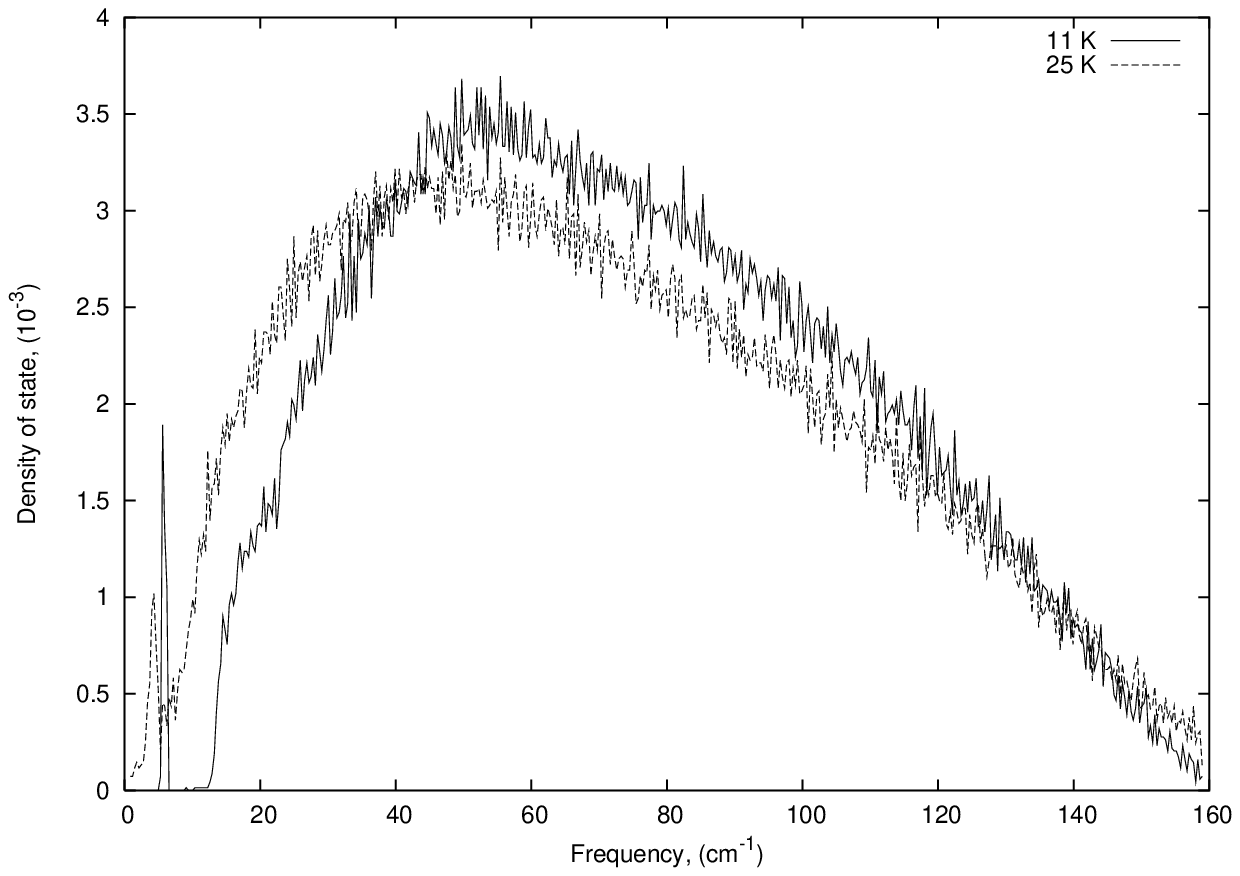}
 \\
 \parbox[t]{0.48\textwidth}{\caption{The DOS for a 229 $CH_4$ cluster, a {\it fcc}-starting configuration at 11 {\it K} - the total spectrum is
indicated with 'all' in the figure inlet; 'rot'  is the rotational and 'vib' is the vibrational spectrum.}\label{vra-spec}}
 \hfill
 \parbox[t]{0.48\textwidth}{\caption{Vibrational spectrum for 229 $CH_4$ cluster; {\it fcc} - start.}\label{vib-spec}}
\end{figure}

Finally, we present the Density of States (DOS) function 
obtained by a normal mode analysis of the 
quenched system \cite{Weber}.
 The Fig.\ref{vra-spec} shows evidently the existence of a
rovibrational coupling \cite{Bell}. The Fig.\ref{vib-spec}
 shows broadening of the vibrational 
spectrum at heating.
\section{Conclusion}
Implementing Molecular Dynamics method to a microcanonical state
of free methane clusters, we show that they exhibit transformations
between an orientationally-ordered (at $\sim$ 10K) and disordered phase
(above 20K). The lack of hysteresis in the cooling-heating regime indicates
a continuous transition. However, such a statement can be confirmed
either by studying infinite systems (i.e. periodic
boundary conditions) or by applying the finite-size scaling
theory \cite{Fisher}.

\section*{Acknowledgments}
The authors would like to thank Prof. Arias' group from the Cornell University 
for the access to the computer facilities. The work is supported by a special grant of the Bulgarian Ministry of Education and Science (F-3, 2003).

\end{document}